\documentclass[article,twocolumn,aps,prl,groupedaddress,showpacs]{revtex4-2}

\usepackage{amsmath}
\usepackage{amssymb}
\usepackage{txfonts}

\usepackage{gensymb}
\usepackage{graphicx,bm,units,yfonts}
\usepackage{float}
\usepackage[table]{xcolor}

\usepackage{hyperref}

\usepackage{cancel}

\usepackage[T1]{fontenc}
\usepackage{lmodern}

\usepackage{multirow}

\begin{document}

\title{
Flat Midgap Topological Surface and Hypersurface Bands without Parameter Tuning
%Flat Midgap Topological Surface and Hypersurface Bands in Three and Higher Dimensional Su-Schrieffer-Heeger Model
}

\author{Keun Hyuk Ahn}
\email{kenahn@njit.edu}
\affiliation{Department of Physics, New Jersey Institute of Technology, Newark, New Jersey 07102, USA}

\begin{abstract}
% 600 character limit
% 500 words limit
The Su-Schrieffer-Heeger model is extended to the three and higher dimensional systems.
Nearly or absolutely flat midgap surface and hypersurface bands are predicted based on the topological analysis,
which do not require fine tuning of parameters.
By adding the on-site Coulomb interaction for the three dimensional systems,
we computationally show that
the large difference in the band widths between the surface and the bulk
leads to the strongly correlated phenomena,
specifically magnetism, confined only on the surface.
Possible experimental realizations in solid state materials and metamaterials 
are discussed.
\end{abstract}

\maketitle

There has been a great interest recently in flat energy or frequency bands 
in solid state materials and metamaterials~\cite{Chalker10, Leykam18b, Rhim21, Bergholtz13, Checkelsky24, Qiu16, Slot17, Jeon22}, 
such as twisted bilayer graphenes~\cite{Bistritzer11, Cao18-Nature-a,Cao18-Nature-b, Andrei20, Torma22} 
and the Lieb photonic crystals~\cite{Lieb89, Wiersma15, Mukherjee15,Vicencio15}. 
Flat bands could result in strongly correlated phenomena, 
such as magnetism and superconductivity, 
or localized excitations, such as trapped light and sound, 
which have both scientific and technological significance. 
However, most of these flat bands are only approximately flat, 
require fine tuning of parameters, touch or cross dispersive bands, 
involve spin-orbit interactions, exist as one-dimensional bands, 
or are hard to realize in solid state materials, 
which limit the realization of the full potentials of the flat bands~\cite{Leykam18b}.

%Flat band systems have attracted great research interest 
%lately~\cite{Chalker10, Leykam18b, Rhim21, Bergholtz13, Checkelsky24, Qiu16, Slot17, Jeon22}.
%Enhanced electron correlation effects in the narrow bands  
%due to the reduced kinetic energy 
%are believed to be responsible for 
%the unconventional superconductivity
%in twisted bilayer graphenes~\cite{Bistritzer11, Cao18-Nature-a,Cao18-Nature-b, Andrei20, Torma22}.
%Localized light propagation observed in the Lieb photonic crystals
%with flat frequency bands 
%highlights 
%the technological potentials of the flat bands 
%in metamaterials~\cite{Lieb89, Wiersma15, Mukherjee15,Vicencio15}.
%The author and collaborators have recently shown theoretically that 
%the topological characteristics of the  
%two-dimensional (2D) Su-Schrieffer-Heeger (SSH) model
%give rise to the flat one-dimensional (1D) edge bands~\cite{Zhu19},
%and experimentally demonstrated them
%using mechanical systems~\cite{Qian20}. 
%However, strongly correlated phenomena
%in flat electronic band systems, such as magnetism and unconventional superconductivity,
%usually require flat bands of two or 
%higher dimensions~\cite{Mermin66, Hohenberg67, Pitaevskii91, Halperin19}.  
%Therefore, the increase of  
%the dimension of the flat bands 
%from 1D to 2D would be a significant advancement.
%This would open up the possibility
%of realizing strongly correlated phenomena
%confined only on the surfaces
%of electronic metamaterials,
%which would provide 
%greater controllability of these phenomena
%than conventional solid state flat band systems~\cite{Cao18-Nature-a,Cao18-Nature-b}. 

In this Letter, we propose nearly or absolutely flat, 
midgap surface and hypersurface bands 
of topological origins without fine tuning of parameters or without invoking spin-orbit interactions, 
which are likely to be realized in solid state materials 
and metamaterials. 
This is achieved 
by extending the prior works by the author and collaborators~\cite{Zhu19,Qian20}, that is,
by extending the Su-Shrieffer-Heeger (SSH) model 
to three and higher dimensions~\cite{Price22, Ozawa19, Wang20}, 
applying topological analyses, and computationally demonstrating that the model could 
host flat topological surface bands. 
With the Coulomb interaction considered, we further demonstrates magnetic orderings confined only on the surfaces, 
as an example of surface-confined strongly correlated phenomena.
Discussion is provided on how such flat surface and hypersurface bands
could be realized experimentally.

%In this Letter, we propose a three-dimensional (3D) chiral extension 
%of the SSH model~\cite{Asboth16-Book},
%show the presence of the 2D flat midgap 
%surface bands of the topological origin,
%and numerically demonstrate
%that the Coulomb interaction between electrons 
%could result in strongly correlated phenomena, 
%specifically magnetic ordering, only on the surfaces.
%We also study the systems 
%with one of the two parallel open surfaces made of half unit cells, and reveal 
%a completely flat band confined only on one surface,
%for which any nonzero Coulomb interactions 
%would polarize the spins on the surface.
%The extension to higher $N$-dimensions ($N$D)
%is presented~\cite{Price22, Ozawa19, Wang20} 
%and possible experimental realizations 
%of both 3D and higher $N$D SSH models are discussed.

\begin{figure}
\includegraphics[width=0.8\linewidth]{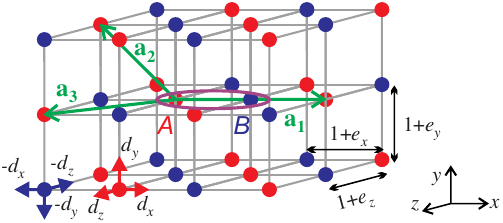}
\caption{
3D chiral SSH system. 
The purple ellipse represents the unit cell made of two sites, $A$ and $B$,
shown in red and blue dots. 
The green arrows show the primitive vectors, ${\bf a}_1$, ${\bf a}_2$, and ${\bf a}_3$,
and the red and blue arrows the staggered distortions parameterized by 
$d_x$, $d_y$, and $d_z$ in the $x$, $y$, and $z$ directions. 
$1+e_x$, $1+e_y$, and $1+e_z$ represent the intersite distances 
before the staggered distortions are introduced.
}
\label{fig-structure}
\end{figure}

Starting with a simple cubic lattice with a single-site basis and a unit lattice constant,
we add uniform strains, $e_x$, $e_y$, and $e_z$, in the $x$, $y$, and $z$ directions,
which change the lattice constants to $1+e_x$, $1+e_y$, and $1+e_z$.
Further, 3D checkerboard style staggered distortions,
parameterized by $d_x$, $d_y$, and $d_z$,
are added as shown in Fig.~\ref{fig-structure}, 
which doubles the unit cell, marked by the purple ellipse, 
and results in a basis of two sites, $A$ and $B$.
Primitive vectors
${\bf a}_1$, ${\bf a}_2$, and ${\bf a}_3$ 
are chosen as shown in Fig.~\ref{fig-structure}. 

With the nearest-neighbor electron hopping parameters 
modulated by the variations in the intersite distances, 
the tight binding Hamiltonian of the system 
without surfaces is given by
\begin{eqnarray}
&&\hat{H}_0=\sum_{{\bf R},\sigma=\uparrow,\downarrow} 
-(\alpha_x+\beta_x) c^{\dagger}_{{\bf R} B \sigma} c_{{\bf R} A \sigma} \nonumber \\
&&-(\alpha_x-\beta_x) c^{\dagger}_{{\bf R}+{\bf a}_2, B \sigma} c_{{\bf R} A \sigma} 
-(\alpha_y+\beta_y) c^{\dagger}_{{\bf R}+{\bf a}_3, B \sigma} c_{{\bf R} A \sigma} \nonumber \\ 
&&-(\alpha_y-\beta_y) c^{\dagger}_{{\bf R} B \sigma} c_{{\bf R}+{\bf a}_1, A \sigma} 
-(\alpha_z+\beta_z) c^{\dagger}_{{\bf R} B \sigma} c_{{\bf R}+{\bf a}_1+{\bf a}_2, A \sigma} \nonumber \\
&&-(\alpha_y-\beta_y) c^{\dagger}_{{\bf R} B \sigma} c_{{\bf R}+{\bf a}_1+{\bf a}_3, A \sigma} 
+ {\rm H.c.},
\label{eq-H0-real}
\end{eqnarray}
where 
${\bf R}$ represents the Bravais lattice points,
$c^{\dagger}_{{\bf R}, A/B, \sigma}$ is
the creation operator for
the electron with spin $\sigma$ = $\uparrow$, $\downarrow$
at the $A$ or $B$ site within the unit cell at ${\bf R}$,
and the electron hopping amplitudes are parameterized by 
$\alpha_a=t_0(1-e_a)$ and
$\beta_a=2 t_0 d_a$ 
with $a$=$x$, $y$, $z$,
and $t_0$ representing the nearest-neighbor hopping amplitude 
for the cubic lattice before the distortions.
The Hamiltonian has the chiral symmetry~\cite{Asboth16-Book}, 
as there is no hopping between the $A$ and $A$
or $B$ and $B$ sites,
which gives rise to energy levels 
symmetric about the zero energy.  
With 
$c^{\dagger}_{{\bf R},A/B, \sigma}=\frac{1}{\sqrt N_{\rm lattice}} \sum_{\bf k} c^{\dagger}_{{\bf k}, A/B, \sigma} e^{-i \bf{k} \cdot \bf{R}}$,
${\bf k}= (k_1/2\pi) {\bf b}_1 + (k_2/2\pi) {\bf b}_2 + (k_3/2\pi) {\bf b}_3$,
${\bf b}_n$ ($n$ = 1, 2, 3) representing primitive reciprocal vectors,
and $N_{\rm lattice}$ the number of the Bravais lattice points, 
the Hamiltonian is represented in the form of 
\begin{equation}
\hat{H}_0=\sum_{\bf k, \sigma=\uparrow,\downarrow}
\left[h^{*}({\bf k}) c^{\dagger}_{{\bf k} A \sigma}  
c_{{\bf k} B \sigma} +
h ({\bf k}) c^{\dagger}_{{\bf k} B \sigma}
c_{{\bf k} A \sigma}\right],
\end{equation}
which leads to the spin-degenerate energy bands of
$\varepsilon_{0,\pm}({\bf k})=\pm | h ({\bf k}) |$ (see Supplemental Materials~\cite{SupMat}).

The topological analysis is carried out in terms of 
three winding numbers, defined by
\begin{equation}
\nu_n=\frac{1}{2\pi i} \int^{2 \pi}_0 dk_n \frac{\partial}{\partial k_n} \ln h(k_1,k_2,k_3),
\end{equation} 
with $n$ = 1, 2, 3~\cite{Asboth16-Book, Zhu19, Delplace11, Zak89}.
The results are shown in Table~\ref{table-winding} for chosen conditions,
in which the systems with any nonzero winding numbers are 
topological insulators (TI) and would host 
flat midgap bands on the open surfaces perpendicular 
to the primitive reciprocal vectors associated with the nonzero winding numbers.  
We first focus on the top two rows in Table~\ref{table-winding},
in which the condition of $\beta_x$$\neq$0 and $\beta_y$=$\beta_z$=0 
describes the SSH chains running in the $x$ direction.
The condition of $\alpha_y$+$\alpha_z$<$\alpha_x$ 
means 
that the sum of the {\it interchain} couplings should be weaker 
than the average {\it intrachain} coupling,
which is the condition for the opening of the band gap 
and an insulating phase for a half filling. 
Depending on the sign of the staggered distortion $\beta_x$ 
the system could be either topological or nontopological insulators.
A similar interpretation applies to all other cases in Table~\ref{table-winding},
which represent the weakly coupled SSH chains running in $y$ or $z$ direction.
There are two types of topological surfaces.
One type occurs for the systems with $\nu_1$$\neq$0, 
in which the topological surfaces are  
perpendicular to ${\bf b}_1$, the body-diagonal,
or parallel to the plane defined by ${\bf a}_2$ and ${\bf a}_3$,
and have either $A$ sites only or $B$ sites only 
on each open surface and in each layer parallel to it.
The other type occurs for the systems with $\nu_2$$\neq$0 or $\nu_3$$\neq$0,
in which the topological surfaces are perpendicular to
the SSH chains.
In this Letter, we focus on the first type to limit the scope. 

\begin{table} 
\begin{tabular} {c | c | c c c | c c c | c}
\hline \hline
SSH chains &
$\alpha_x$, $\alpha_y$, $\alpha_z$ & $\beta_x$ & $\beta_y$ & $\beta_z$ & $\nu_1$ & $\nu_2$ & $\nu_3$ & TI/nonTI  \\
\hline
\multirow{2}{*}{Along $\hat{x}$} & \multirow{2}{*}{$\alpha_y$+$\alpha_z$<$\alpha_x$} & $+$ & 0 & 0 & 0 & 0 & 0 & nonTI \\
& & $-$ & 0 & 0 & 1 & 0 & 0 & TI \\
\hline
\multirow{2}{*}{Along $\hat{y}$} & \multirow{2}{*}{$\alpha_z$+$\alpha_x$<$\alpha_y$} & 0 & $+$ & 0 & 0 & -1 & 0 & TI \\
& & 0 & $-$ & 0 & 1 & 1 & 0 & TI \\
\hline
\multirow{2}{*}{Along $\hat{z}$} & \multirow{2}{*}{$\alpha_x$+$\alpha_y$<$\alpha_z$} & 0 & $0$ & $+$ & 0 & 0 & -1 & TI \\
& & 0 & 0 & $-$ & 1 & 0 & 1 & TI \\
\hline
\end{tabular}
\caption{
Winding numbers $\nu_n$ ($n$=1, 2, 3)
for surfaces parallel to 
the primitive vectors ${\bf a}_m$ and ${\bf a}_{m'}$($m$, $m'\neq n$)
for the 3D SSH system shown in Fig.~\ref{fig-structure}.
The parameters 
$\alpha_a$ and $\beta_a$ ($a$=$x$, $y$, $z$)
are the parameters in the Hamiltonian in Eq.~(\ref{eq-H0-real}), 
associated with the distortions.
TI means topological insulator.   
}
\label{table-winding}
\end{table}

\begin{figure}
\includegraphics[width=0.9\linewidth]{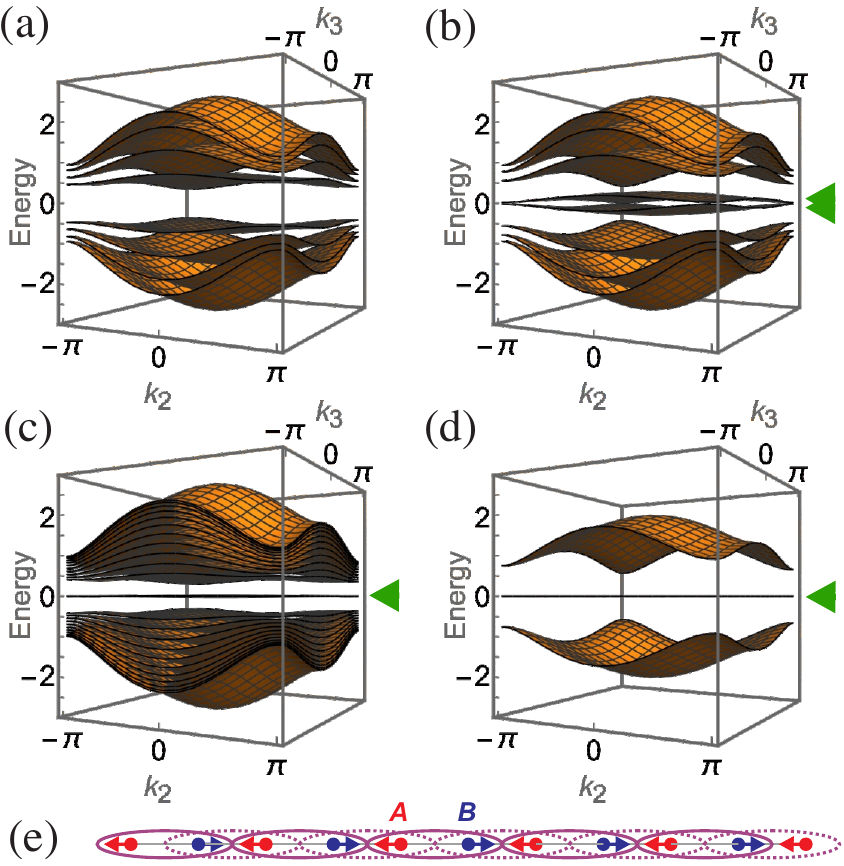}
\caption{
(a-d): Band structures for the 3D SSH systems with open surfaces
before the Hubbard term is considered.
(e): 1D SSH chain with a half unit cell on one edge.
(a-c) show the results for systems with full unit cells on both surfaces,
while (d) for a system with half unit cells on one surface.
The green triangles mark the surface bands.
Common parameter values are $\alpha_x$=1.0, $\alpha_y$=0.3, $\alpha_z$=0.2,
$\beta_y$=$\beta_z$=0,
while varied parameter values of $N_1$ and $\beta_x$ 
are 4 and 0.2 for (a),
4 and -0.2 for (b),
12 and -0.2 for (c),
and 
1 and $\pm0.2$ for (d). 
In the 1D SSH system in (e),
the unit cells 
chosen from the left and right edges,
marked by ellipses in solid and dotted lines, respectively,
indicate that one edge (the left edge for the shown distortions) is topological 
and the other nontopological~\cite{CP, Benito16}.
}
\label{fig-band-wo-U}
\end{figure}

In electronic band structures, 
the band width reflects the energy scale for the kinetic energy.
Therefore, if the band width is narrow, 
the effects of the Coulomb interaction would be enhanced.
In the 3D topological SSH systems here,
the surface band width is much narrower than the bulk band width, 
which means that the same Coulomb interaction 
would have strikingly different effects between the surface and the bulk,
and could give rise to strongly correlated phenomena only on the surfaces.
To study such effects, 
the on-site electron-electron Coulomb interaction of $U$ is added to the Hamiltonian in Eq.~(\ref{eq-H0-real}),
which leads to the Hubbard-SSH Hamiltonian 
$\hat{H}$=$\hat{H}_0$+$\hat{H}_{\rm C}$ with
$\hat{H}_{\rm C}$=$\sum_{{\bf R}} U \hat{\rho}_{{\bf R}A\uparrow}\hat{\rho}_{{\bf R}A\downarrow}$ 
+$U \hat{\rho}_{{\bf R}B\uparrow}\hat{\rho}_{{\bf R}B\downarrow}$,
representing the on-site Coulomb interaction,
and $\hat{\rho}_{{\bf R},A/B,\sigma}$=$c^{\dagger}_{{\bf R},A/B,\sigma} c_{{\bf R},A/B,\sigma}$
the number operator.
Within the Hartree-Fock approximation, 
the Hamiltonian is transformed to $\hat{H}_{\rm HF}$ shown below, which is solved recursively 
till the self-consistency is reached.
\begin{eqnarray}
\hat{H}_{\rm HF}&=&\hat{H}_0+\hat{H}_{\rm C,HF} \\
\hat{H}_{\rm C,HF}&=&\sum_{{\bf R}} 
U \hat{\rho}_{{\bf R}A\uparrow}<\hat{\rho}_{{\bf R}A\downarrow}> 
+U <\hat{\rho}_{{\bf R}A\uparrow}>\hat{\rho}_{{\bf R}A\downarrow} \nonumber \\
& &+ U \hat{\rho}_{{\bf R}B\uparrow}<\hat{\rho}_{{\bf R}B\downarrow}> 
+ U <\hat{\rho}_{{\bf R}B\uparrow}>\hat{\rho}_{{\bf R}B\downarrow}
\end{eqnarray} 
Various periodicities are considered
and the lowest energy states are found
for the cases with $\beta_x \neq 0$
in Table~\ref{table-winding}. 
We first study the systems without any surfaces.
To be specific, the parameter values are chosen as shown 
in the caption of Fig.~\ref{fig-band-wo-U},
and we seek trends in numerical results.
For the $U$=0 case, the ground state is nonmagnetic 
and, for the chosen parameter values,
the width of each band is about 2.7 
and the size of the gap about 0.7.
As the Coulomb interaction $U$ increases past $U_c$$\approx$2.0, the same order of magnitude as the band width, 
the ground state changes to a 3D checkerboard-type antiferromagnetic state 
with $A$ and $B$ sites having the opposite spin directions.

Numerical approaches are used to study the systems with the open surfaces
parallel to ${\bf a}_2$ and ${\bf a}_3$. 
The total number of layers is $L_1$=$2N_1$ if both surfaces are made of full unit cells,
and $L_1$=$2N_1$+1 if one surface is made of half unit cells,
where $N_1$ represents the number of full unit cells along the ${\bf a}_1$ direction.
Figure~\ref{fig-band-wo-U} shows the band structures
for the systems with open surfaces,
before the Coulomb term, $\hat{H}_{\rm C, HF}$, is included.
The band structures for the nonTI and TI systems are 
shown in Figs.~\ref{fig-band-wo-U}(a) and \ref{fig-band-wo-U}(b), 
respectively,
with $N_1$=4 and full unit cells on both surfaces,
which reveal clear differences.
While the nonTI band structure in Fig.~\ref{fig-band-wo-U}(a)
has no bands in the gap, 
the TI band structure in Fig.~\ref{fig-band-wo-U}(b)
has two surface bands in the gap, marked by the green triangles,
consistent with the topological analysis.
The small dispersions (about 0.28) for each surface band 
and the tiny separation (about 0.04) between them
in Fig.~\ref{fig-band-wo-U}(b)
are the results of the hybridization of the surface bands from the opposite surfaces,
similar to the 1D and 2D SSH systems~\cite{Asboth16-Book,Zhu19,Qian20}.
One way to reduce the hybridization is 
to increase the separation between the open surfaces,
as demonstrated by the band structures for a thicker, $N_1$=12, TI system 
in Fig.~\ref{fig-band-wo-U}(c). 
The two midgap surface bands are indeed almost completely flat 
(the width of about 0.03)
and their separation is greatly reduced (down to $4 \times 10^{-5}$).
Another way to achieve flatter, in fact {\it absolutely} flat, surface bands is 
to eliminate the surface band on the other surface by making it 
locally equivalent to the nontopological surface.
This is achieved by having half unit cells on one surface.
In this case, 
analogous to the 1D SSH system shown in Fig.~\ref{fig-band-wo-U}(e)~\cite{CP, Benito16}, 
the unique characteristics of the SSH chain make
one surface topological 
and the other nontopological
regardless of the sign of the 
staggered distortion $\beta_x$, 
and the topological midgap surface band is present only on one surface and absolutely flat. 
The band structure in Fig.~\ref{fig-band-wo-U}(d)
 for $N_1$=1 with $\beta_x$$\neq$0 shows that,
 unlike the full unit cell TI in Fig.~\ref{fig-band-wo-U}(b), 
 only one spin-degenerate surface band exists and is completely flat 
 with energy zero~\cite{SupMat}.
The 3D SSH systems here, regardless of full or half unit cells 
on the surfaces, have the chiral symmetry,
which guarantees that
the eigenstates appear in pairs with the energies 
$\pm\varepsilon({\bf k}_{\parallel})$ at each wavevector ${\bf k}_{\parallel}$ 
parallel to the surface~\cite{Asboth16-Book}.
If the number of layers is odd,
the chiral symmetry forces at least one eigenstate to have a zero energy at each ${\bf k}_{\parallel}$,
and the system has at least one completely flat band,
as Fig.~\ref{fig-band-wo-U}(d) shows.  

\begin{figure}
\includegraphics[width=\linewidth]{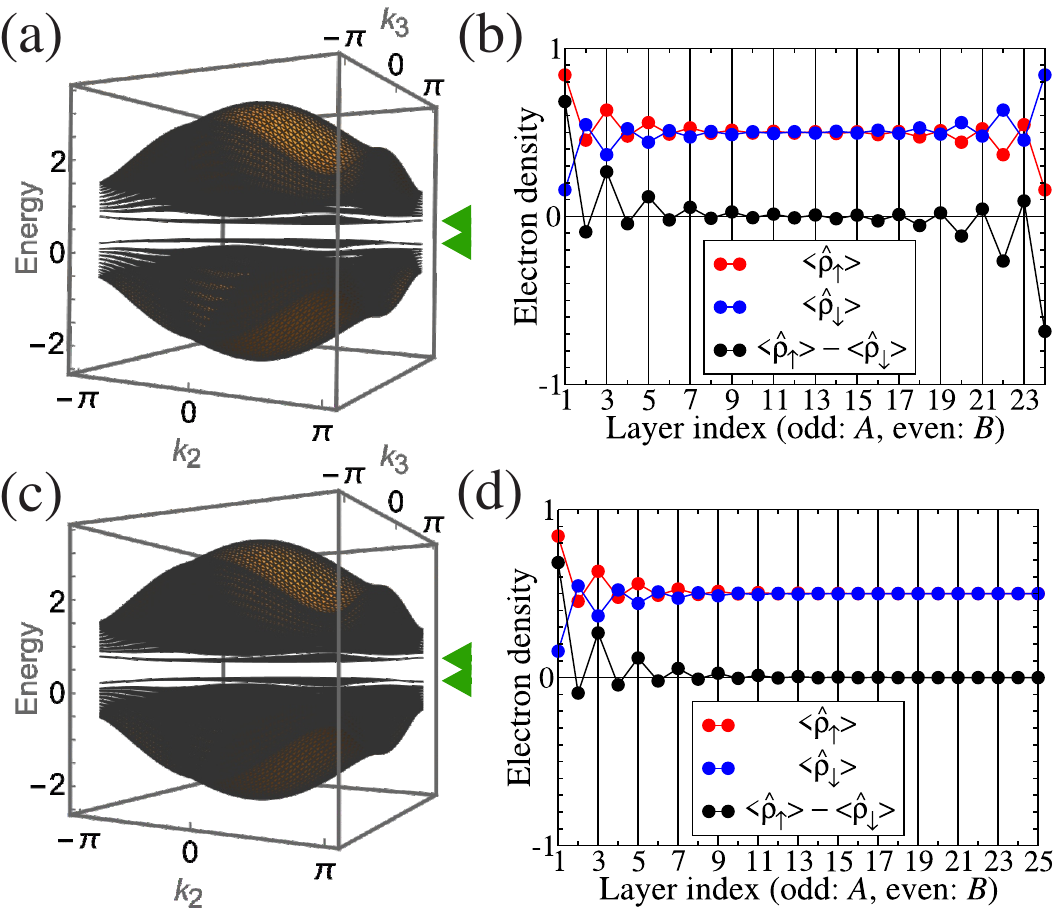}
\caption{
Results for the 3D Hubbard SSH systems with open surfaces
within the Hartree-Fock approximations.
(a), (b): for the system with full unit cells on both surfaces. 
(c), (d): for the system with half unit cells on one surface.
(a), (c): the band structures with green triangles marking the surface bands. 
(b), (d): the profiles for the
spin up and spin down electron densities and their difference, that is,
<$\hat{\rho}_{\uparrow}$>,
<$\hat{\rho}_{\downarrow}$>,
and <$\hat{\rho}_{\uparrow}$>-<$\hat{\rho}_{\downarrow}$>, 
versus the index of layers.
The Coulomb interaction is $U$=1.0,
and other parameter values are identical to those for Fig.~\ref{fig-band-wo-U}(c). 
}
\label{fig-band-map-w-U}
\end{figure}

The striking band width difference 
between the surface and the bulk
seen here for the TI systems 
provides the opportunity to create strongly correlated phases
only on the surfaces, not in the bulk.
We choose the on-site Coulomb interaction $U$=1.0,
greater than the surface band width but less than the bulk band width,
and find the ground state.
The results are shown in Fig.~\ref{fig-band-map-w-U}. 
Figure~\ref{fig-band-map-w-U}(a) shows 
the spin degenerate band structure
for the TI systems with full unit cells on both surfaces and $N_1$=12. 
Comparison with Fig.~\ref{fig-band-wo-U}(c)
reveals that
the two surface bands, 
marked by the green triangles in Fig.~\ref{fig-band-map-w-U}(a),
split by about $U$, 
while each surface band remaining quite narrow.
The spin up and spin down electron densities,
represented by <$\hat{\rho}_{\uparrow}$> and<$\hat{\rho}_{\downarrow}$>,
are found to be constant in the ${\bf a}_2$ and ${\bf a}_3$ directions,
but modulating along the ${\bf a}_1$ direction
while maintaining the total electron density constant, that is, 
 <$\hat{\rho}_{\uparrow}$>+<$\hat{\rho}_{\downarrow}$>=1.
Figure~\ref{fig-band-map-w-U}(b) shows 
the profiles of the spin up and spin down electron densities 
and their difference along the ${\bf a}_1$ direction
in red, blue, and black symbols, respectively.
The results show that the net magnetic moment (black symbols)
has the greatest magnitude at the open surfaces,
and decays rapidly towards the interior
with the sign alternating between the $A$ and $B$ layers,
demonstrating different effects of the Coulomb interaction on the surfaces and the bulk. 

The electron density profiles for the system with half unit cells 
on one surface for the same $U$=1.0, shown in Fig.~\ref{fig-band-map-w-U}(d),
is quite different.
The net magnetic moment (black symbols) appears only on one surface,
but is absent on the other,
as only one surface hosts the flat topological surface band.
Although its band structure in Fig.~\ref{fig-band-map-w-U}(c)
may look overall similar to that in Fig.~\ref{fig-band-map-w-U}(a),
the two flat surface bands, marked by green triangles, 
have opposite spins
and all the surface and bulk bands are nondegenerate,
as the electron density profile 
in Fig.~\ref{fig-band-map-w-U}(d) breaks the global spin symmetry. 

\begin{figure}
\includegraphics[width=\linewidth]{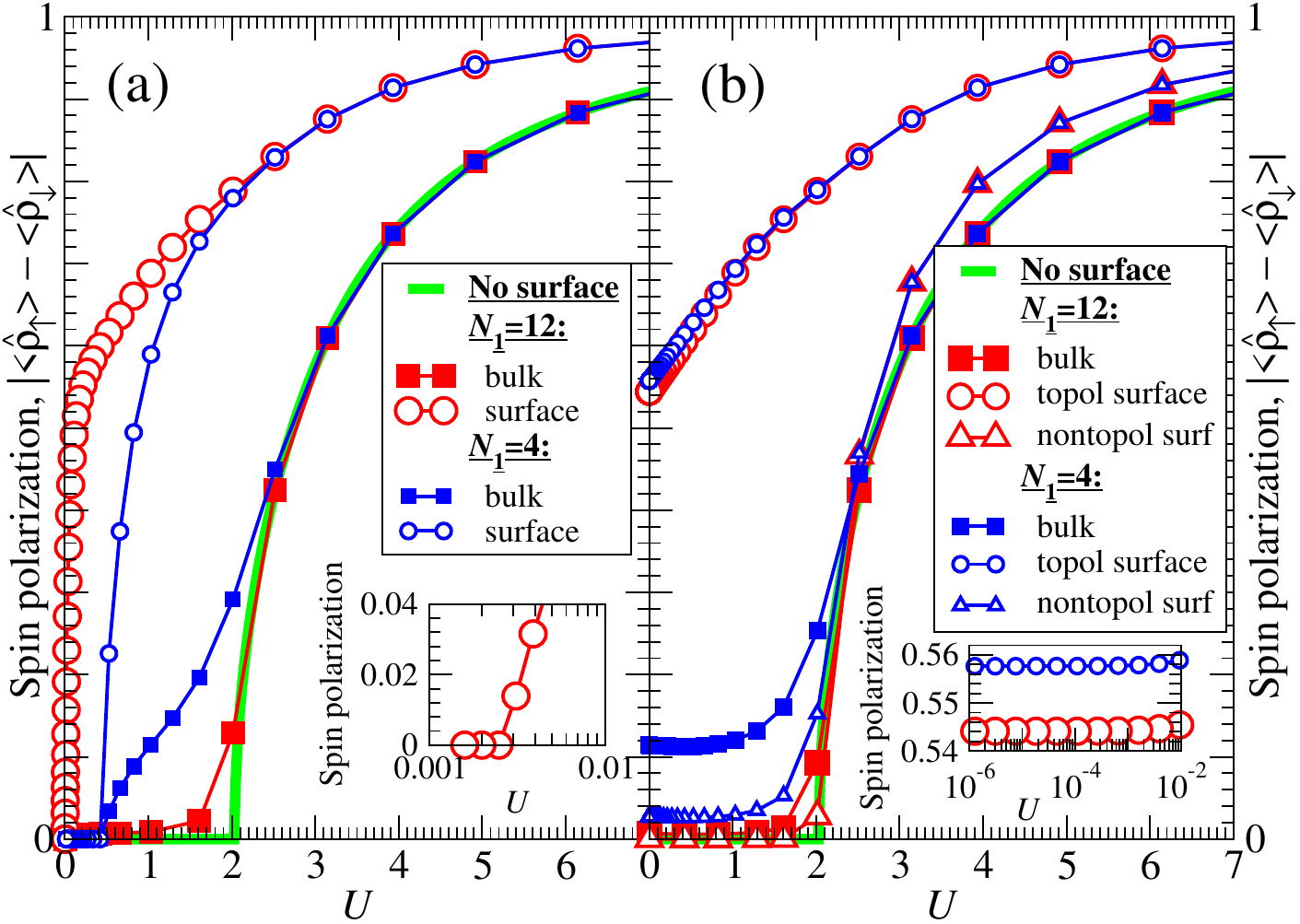}
\caption{
Surface and bulk spin polarizations, |<$\hat{\rho}_{\uparrow}$>-<$\hat{\rho}_{\downarrow}$>|,
versus the Coulomb interaction $U$ 
for the 3D Hubbard SSH systems.
(a) for the systems with full unit cells on both surfaces,  
and (b) for the systems with half unit cells on one surface.
The thick systems with $N_1$=12 (red symbols) and 
the thin systems with $N_1$=4 (blue symbols) are considered.  
Other parameter values are identical to those 
in Figs.~\ref{fig-band-wo-U}(c) and ~\ref{fig-band-map-w-U}.
The green lines show the results for the systems without open surfaces.
The circles and squares represent
the spin polarization on the topological surfaces 
and at the center layer in the bulk, respectively.
The triangles in (b) represent 
the spin polarization on nontopological surfaces.
The insets show the spin polarization at small $U$'s in semilogarithmic scales.
}
\label{fig-M-v-U}
\end{figure}

The large difference between the surface and the bulk band widths
seen in Fig.~\ref{fig-band-wo-U}
manifests itself 
as very different critical Coulomb interactions for the magnetizations
of the surface and the bulk.
The spin polarization, defined as |<$\hat{\rho}_{\uparrow}$>-<$\hat{\rho}_{\downarrow}$>|,
versus the Coulomb interaction $U$
for the systems with full unit cells on both surfaces
is displayed in Fig.~\ref{fig-M-v-U}(a).
The result for the system without surfaces
is also shown in green solid line. 
For the thin system with $N_1$=4 (blue symbols),
the surface spin polarization (blue circles)
starts to appear at around $U_{\rm c,s}$$\approx$0.42,
the same order of magnitude as the width  
of each surface band (about 0.28) in Fig.~\ref{fig-band-wo-U}(b),
and increases rapidly, approaching to the full spin polarization of 1.0.
As for the bulk spin polarization (blue squares), 
defined as |<$\hat{\rho}_{\uparrow}$>-<$\hat{\rho}_{\downarrow}$>| at the center layer,
although it appears at the same $U$$\approx$$U_{\rm c,s}$$\approx$0.42 
as the surface spin polarization
and increases slowly, 
this is the tail of the surface spin polarization 
seen in Fig.~\ref{fig-band-map-w-U}(b),
and the {\it intrinsic} bulk spin polarization develops rapidly  
only above $U_{\rm c,b}\approx$2.0, 
same as $U_{\rm c}$ for the system without surfaces (see the green line).
As the system becomes thicker to $N_1$=12 (red symbols),
the surface band becomes much narrower, 
as seen in Fig.~\ref{fig-band-wo-U}(c),
and the surface spin polarization (red circles) appears 
from around $U_{\rm c,s}\approx$0.0025 as revealed in the inset,
about 200 times smaller than that of $N_1$=4 system.
The tails of the surface states are suppressed substantially
at the center layer
and the bulk spin polarization (red squares) remains negligible
until the intrinsic bulk spin polarization develops 
at around $U_{\rm c,b}$$\approx$2.0. 
Further studies on the systems with $N_1$=6, 8, 10 reveal 
that the decay in $U_{\rm c,s}$ with respect to $N_1$ is exponential,
reflecting the exponentially decaying surface band width~\cite{Qian20} 
(see Supplemental Materials for the phase diagrams~\cite{SupMat}).

The spin polarization versus
the Coulomb interaction
for the systems with one surface 
made of half unit cells
is shown in Fig.~\ref{fig-M-v-U}(b).
Because the surface band for $U$=0 
is absolutely flat,
as seen in Fig.~\ref{fig-band-wo-U}(d),
any nonzero Coulomb interaction $U$ 
gives rise to the magnetization
on the topological surface,
as shown in Fig.~\ref{fig-M-v-U}(b) and its inset 
for 
$N_1$=4 (blue circles) 
and
$N_1$=12 (red circles).
For the thin $N_1$=4 systems, 
the tails of the surface states
show up as finite spin polarizations
in the center layer of the bulk (blue squares) and 
even on the nontopological surface (blue triangles).
Similar to Fig.~\ref{fig-M-v-U}(a),
such spin polarizations decrease rapidly 
when the system thickness increases to 
$N_1$=12 (red squares and red triangles).
   
By making artificial connections,
metamaterials with effective dimensions 
higher than three could be created~\cite{Ozawa19, Price22, Wang20}.
Therefore, the extension of the 3D SSH model to 
higher $N$D is of interest.
As in the 3D SSH models, 
the $N$D chiral SSH model 
is made of 1D SSH chains
coupled in the $N$-1 dimensional directions, 
with neighboring chains shifted with respect to each other
and the $A$ sites surrounded by the $B$ sites
and vice versa.
These 1D SSH chains should couple 
weakly to open a band gap.
Specifically, the sum of the {\it interchain} couplings 
should be weaker than the average {\it intrachain} coupling.  
With the primitive vectors of the $N$D SSH system defined 
in the way analogous to those in Fig.~\ref{fig-structure},  
the $N$ winding numbers are defined.
Just like the 3D SSH systems, 
there could be two kinds of topological hypersurfaces.
The first is perpendicular to the $N$D diagonal direction,
and the second perpendicular to the SSH chains.
Further details, including $N$D SSH Hamiltonian, energy bands, winding numbers, topological analysis,
and nearly or absolutely flat, topological, midgap hypersurface bands, 
are given in Supplemental Materials~\cite{SupMat}.  
   
For the 3D SSH systems, 
it would be fruitful for the future 
studies to investigate the possibility 
of other strongly correlated phenomena,
particularly, 
unconventional superconductivity~\cite{Iglovikov14, Peri21},
similar to that 
in twisted bilayer graphenes~\cite{Cao18-Nature-a,Cao18-Nature-b}.
Realizing 3D SSH systems experimentally
would require either building artificial structures~\cite{Belopolski17},
such as 3D quantum dot arrays~\cite{Reimann02, Tamura02, Tamura03, Benito16, Piquero-Zulaica17, Kuo21},
or solid state materials search using first-principles calculations and topological indices, as done by Jeon and Kim~\cite{Jeon22}.
The higher $N$D SSH systems
would be realized by building multiple 3D systems 
and connecting them using electronics
that mimic the higher dimensional interactions~\cite{Price22, Ozawa19, Wang20}.

In summary, we have proposed the chiral extension of the SSH model 
to three dimensions,
and showed that it could 
give rise to topological insulators
with flat midgap surface bands.
It is demonstrated that 
the Coulomb interaction could result in magnetism
confined only on the topological surfaces.
The extension to higher dimensions is presented.


\begin{thebibliography}{}

% Flat band general

\bibitem{Chalker10}
J. T. Chalker, T. S. Pickles, and P. Shukla, 
Anderson localization in tight-binding models with flat bands,
Phys. Rev. B {\bf 82}, 104209 (2010).

\bibitem{Bergholtz13}
E. J. Bergholtz and Z. Liu, 
Topological flat band models and fractional Chern insulators,
Int. J. Mod. Phys. B {\bf 27}, 1330017 (2013).

\bibitem{Leykam18b}
D. Leykam, A. Andreanov, and S. Flach, 
Artificial flat band systems: From lattice models to experiments, 
Adv. Phys. X {\bf 3}, 1473052 (2018).

\bibitem{Rhim21}
J.-W. Rhim and B.-J. Yang, 
Singular flat bands, 
Adv. Phys. X {\bf 6}, 1 (2021).

\bibitem{Checkelsky24}
J. G. Checkelsky, B. A. Bernevig, P. Coleman, Q. Si and S. Paschen,
Flat bands, strange metals and the Kondo effect,
Nat. Rev. Mater. {\bf 9}, 509 (2024).

\bibitem{Qiu16}
W.-X. Qiu, S. Li, J.-H. Gao, Y. Zhou, and F.-C. Zhang,
Designing an artificial Lieb lattice on a metal surface,
Phys. Rev. B {\bf 94}, 241409(R) (2016).

\bibitem{Slot17}
M. R. Slot, T. S. Gardenier, P. H. Jacobse, G. C. P. van Miert, S. N. Kempkes,
S. J. M. Zevenhuizen, C. M. Smith, D. Vanmaekelbergh, and I. Swart,
Experimental realization and characterization of an electronic Lieb lattice,
Nature Phys. {\bf 13}, 672 (2017).

\bibitem{Jeon22}
S. Jeon and Y. Kim, 
Two-dimensional weak topological insulators in inversion-symmetric crystals,
Phys. Rev. B {\bf 105}, L121101 (2022).

% Twisted bilayer graphene

\bibitem{Bistritzer11}
R. Bistritzer and A. H. MacDonald,
Moir\'{e} bands in twisted double-layer graphene,
Proc. Nat. Acad. Sci. {\bf 108}, 12233 (2011).

\bibitem{Cao18-Nature-a}
Y. Cao, V. Fatemi, S. Fang, K. Watanabe, T. Taniguchi,
E. Kaxiras, and P. Jarillo-Herrero, 
Unconventional superconductivity in magic-angle graphene superlattices,
Nature {\bf 556}, 43 (2018).

\bibitem{Cao18-Nature-b}
Y. Cao, V. Fatemi, A. Demir, S. Fang, S. L. Tomarken, J. Y. Luo, J. D. Sanchez-Yamagishi, K. Watanabe, T. Taniguchi, E. Kaxiras, R. C. Ashoori, and P. Jarillo-Herrero, 
Correlated insulator behaviour at half-filling in magic-angle graphene superlattices,
Nature {\bf 556}, 80 (2018).

\bibitem{Andrei20}
E. Y. Andrei and A. H. MacDonald, 
Graphene bilayers with a twist. 
Nat. Mater. {\bf 19}, 1265 (2020).

\bibitem{Torma22}
P. T\"{o}rm\"{a}, S. Peotta, and B. A.Bernevig, 
Superconductivity, superfluidity and quantum geometry in twisted multilayer systems,
Nat. Rev. Phys. {\bf 4}, 528 (2022).

% Lieb photonic lattice 

\bibitem{Lieb89}
E. H. Lieb, 
Two Theorems on the Hubbard Model, 
Phys. Rev. Lett. {\bf 62}, 1201 (1989).

\bibitem{Mukherjee15}
 S. Mukherjee, A. Spracklen, D. Choudhury, N. Goldman,
P. {\"O}hberg, E. Andersson, and R. R. Thomson, 
Observation of Localized States in Lieb Photonic Lattices, 
Phys. Rev. Lett. {\bf 114}, 245504 (2015).

\bibitem{Vicencio15}
R. A. Vicencio, C. Cantillano, L. Morales-Inostroza,
B.Real, C.Mej{\'\i}a-Cort{\'e}s,S.Weimann,A.Szameit, and
M. I. Molina, 
Observation of a Localized Flat-Band State in a Photonic Lieb Lattice, 
Phys. Rev. Lett. {\bf 114}, 245503 (2015).

\bibitem{Wiersma15}
D. S. Wiersma, 
Trapped in a Photonic Maze,
Physics {\bf 8}, 55 (2015).

\bibitem{Zhu19}
L. Zhu, E. Prodan, and K. H. Ahn, 
Flat energy bands within antiphase and twin boundaries and at open edges in topological materials, 
Phys. Rev. B {\bf 99}, 041117(R) (2019).

\bibitem{Qian20}
K. Qian, L. Zhu, K. H. Ahn, and C. Prodan, 
Observation of flat frequency bands at open edges and antiphase boundary seams in topological mechanical metamaterials, 
Phys. Rev. Lett. {\bf 125}, 225501 (2020).

% Dimensionality and phase transition

%\bibitem{Mermin66}
%N. D. Mermin and H. Wagner,
%Absence of ferromagnetism or antiferromagnetism
%in one-or two-dimensional isotropic Heisenberg models,
%Phys. Rev. Lett. {\bf 17}, 1133 (1966)
%
%\bibitem{Hohenberg67}
%P. C. Hohenberg, 
%Existence of long-range order in one and two dimensions. 
%Phys. Rev. {\bf 158}, 383 (1967)
%
%\bibitem{Pitaevskii91}
%L. Pitaevskii and S. Stringari,
%Uncertainty principle, quantum fluctuations, and broken symmetries.
%J. Low Temp. Phys. {\bf 85}, 377 (1991).
%
%\bibitem{Halperin19}
%B. I. Halperin,
%On the Hohenberg–Mermin–Wagner Theorem and Its Limitations,
%J. of Stat. Phys. {\bf 175} 521 (2019).

% Higher dimensional systems

\bibitem{Ozawa19}
T. Ozawa, H. M. Price, 
Topological quantum matter in synthetic dimensions,
Nat. Rev. Phys. {\bf 1}, 349 (2019).

\bibitem{Price22}
H. Price,
Simulating four-dimensional physics in the laboratory,
Physics Today {\bf 75} (4), 38 (2022).

\bibitem{Wang20}
Y. Wang, H. M. Price B. Zhang, and Y. D. Chong,
Circuit implementation of a four-dimensional topological insulator,
Nat. Commun. {\bf 11}, 2356 (2020). 

% SSH, Topological insulator - General review

\bibitem{Asboth16-Book}
J. K. Asb\'{o}th, L. Oroszl\'{a}ny, and A. P\'{a}lyi, 
{\it A Short Course on Topological Insulators: Band Structure and Edge States in one and Two Dimensions} 
(Springer, Cham, 2016).

% Supplemental Materials

\bibitem{SupMat}
See Supplemental Materials at http://...
for the 3D Hamiltonian and energy bands for the systems without surfaces,
the 3D Hamiltonians for the systems with surfaces, 
the phase diagrams in the plane of the Coulomb interaction versus the system thickness,
and $N$D SSH Hamiltonian and its topological analysis.

% Winding numbers
\bibitem{Delplace11}
P. Delplace, D. Ullmo, and G. Montambaux, 
Zak phase and the existence of edge states in graphene
Phys. Rev. B {\bf 84}, 195452 (2011).

\bibitem{Zak89}
J. Zak, 
Berry’s phase for energy bands in solids,
Phys. Rev. Lett. {\bf 62}, 2747 (1989).

% Half-unit cell surface in spinner and in quantum dot arrays
\bibitem{CP} C. Prodan (private communication)
 
\bibitem{Benito16}
M. Benito, M. Niklas, G. Platero, and S. Kohler,
Edge-state blockade of transport in quantum dot arrays,
Phys. Rev. Lett. {\bf 90}, 067204 (2016).

% Superconductivity in flat band system

\bibitem{Iglovikov14}
V. I. Iglovikov, F. H\'{e}bert, B. Gr\'{e}maud, G. G. Batrouni, and R. T. Scalettar,
Superconducting transitions in flat-band systems,
Phys. Rev. B {\bf 90}, 094506 (2014).

\bibitem{Peri21}
V. Peri, Z.-D. Song, B. A. Bernevig, and S. D. Huber,
Fragile Topology and Flat-Band Superconductivity in the Strong-Coupling Regime,
Phys. Rev. Lett. {\bf 126}, 027002 (2021).

% Quantum dot arrays

\bibitem{Belopolski17}
I. Belopolski, S.-Y. Xu, N. Koirala, C. Liu, G. Bian, V. N. Strocov, G. Chang, M. Neupane,
N. Alidoust, D. Sanchez, H. Zheng, M. Brahlek, V. Rogalev, T. Kim, N. C. Plumb, C. Chen,
F. Bertran, P. Le Fèvre, A. Taleb-Ibrahimi, M.-C. Asensio, M. Shi, H. Lin, M. Hoesch, S. Oh,
M. Z. Hasan, 
A novel artificial condensed matter lattice and a new platform for one-dimensional topological phases, 
Sci. Adv. {\bf 3}, e1501692 (2017).

\bibitem{Reimann02}
S. M. Reimann and M. Manninen,
Electronic structure of quantum dots,
Rev. Mod. Phys. {\bf 74} 1283 (2002).

\bibitem{Tamura02}
H. Tamura, K. Shiraishi, T. Kimura, and H. Takayanagi,
Flat-band ferromagnetism in quantum dot superlattices,
Phys. Rev. B {\bf 65}, 085324 (2002).

\bibitem{Tamura03}
H. Tamura, K. Shiraishi, and H. Takayanagi,
{\it Quantum dot atoms, molecules, and superlattices},
in 
{\it Quantum Dots and Nanowires}, 
edited by S. Bandyopadhyay and H. S. Nalwa (American Scientific Publisher, USA, 2003), Chapter 2.

\bibitem{Piquero-Zulaica17}
I. Piquero-Zulaica, J. Lobo-Checa, A. Sadeghi, Z. M. A. El-Fattah, C. Mitsui, T. Okamoto, R. Pawlak, T. Meier, 
A. Arnau, J. E. Ortega, J. Takeya, S. Goedecker, E. Meyer, and S. Kawai,
Precise engineering of quantum dot array coupling through their barrier widths,
Nat. Commun. {\bf 8}, 787 (2017).

\bibitem{Kuo21}
D. M. T. Kuo, 
High thermoelectric figure of merit of quantum dot array quantum wires,
Jpn. J. Appl. Phys. {\bf 60}, 075001 (2021).

\end{thebibliography}
\end{document}

\title{Supplemental Material for "Flat Midgap Topological Surface and Hypersurface Bands without Parameter Tuning"}

\author{Keun Hyuk Ahn}
\affiliation{Department of Physics, New Jersey Institute of Technology, Newark, New Jersey 07102, USA}

\maketitle

\section{Hamiltonians and energy bands for three-dimensional Su-Schrieffer-Heeger systems}

The off-diagonal element 
$h({\bf k})$ in 
Eq.~(2)
in the main text
is as follows.
\begin{align}
h({\bf k})=&
-(\alpha_x+\beta_x)
-(\alpha_x-\beta_x)e^{ik_1}
-(\alpha_y+\beta_y)e^{-ik_2} \nonumber \\
&-(\alpha_y-\beta_y)e^{i(k_1+k_2)}
-(\alpha_z+\beta_z)e^{-ik_3} \nonumber \\
&-(\alpha_z-\beta_z)e^{i(k_1+k_3)}
\tag{S1}
\end{align}
The bulk energy bands in the absence of the Coulomb interaction 
is given by
\begin{align}
&\varepsilon_{\pm}({\bf k})=\pm 2 \left\{
\left[
\alpha_x \cos{ \frac{k_1}{2} } 
+\alpha_y \cos{ \left(\frac{k_1}{2}+k_2\right) }
\right. \right. \nonumber \\
& \left. \left.
+\alpha_z \cos{ \left(\frac{k_1}{2}+k_3\right) }
\right]^2  \right.
+
\left.
\left[
\beta_x \sin{ \frac{k_1}{2} }
+\beta_y \sin{ \left(\frac{k_1}{2}+k_2\right) }
\right.
\right.
\nonumber \\
&
\left.
\left.
+\beta_z \sin{ \left(\frac{k_1}{2}+k_3\right) }
\right]^2
\right\}^{1/2}.
\tag{S2}
\end{align}

The surface breaks the translational symmetry in one direction,
but maintains it in other directions.
The three-dimensional (3D) Su-Schrieffer-Heeger (SSH) systems with the surfaces 
parallel to ${\bf a}_2$ and ${\bf a}_3$,
the thickness $L_1=2N_1$, 
and the full unit cells on both surfaces 
are represented by the following Hamiltonian in the absence of the Coulomb interaction:
\begin{align}
\hat{H}_0=&\sum_{{\bf k}_{\parallel}, \sigma=\uparrow,\downarrow}
\left(
\begin{array}{c}
C^{\dagger}_{{\bf k}_{\parallel} A \sigma} \\
C^{\dagger}_{{\bf k}_{\parallel} B \sigma}
\end{array}
\right)^T
\nonumber \\
&\left(
\begin{array}{c c}
0 &
h_{N_1 \times N_1}^{\dagger}({\bf k}_{\parallel}) \\
h_{N_1 \times N_1} ({\bf k}_{\parallel}) &
0
\end{array}
\right) 
\left(
\begin{array}{c}
C_{{\bf k}_{\parallel} A \sigma} \\
C_{{\bf k}_{\parallel} B \sigma}
\end{array}
\right),
\tag{S3}
\label{eq-H0-surface-full-unit-cells} 
\end{align}
where
${\bf k}_{\parallel} = (k_2/2\pi) {\bf b}_2 + (k_3/2\pi) {\bf b}_3$,
$C^{\dagger}_{{\bf k}_{\parallel}, A/B, \sigma}$ =
$(c^{\dagger}_{1, {\bf k}_{\parallel},  A/B, \sigma},  
c^{\dagger}_{2, {\bf k}_{\parallel}, A/B, \sigma},  ..., 
c^{\dagger}_{N_1, {\bf k}_{\parallel}, A/B, \sigma})$, and 
the $N_1 \times N_1$ matrix $h_{N_1\times N_1}({\bf k}_{\parallel})$ is
\begin{equation}
\tag{S4}
h_{N_1\times N_1}({\bf k}_{\parallel})=
\left( 
\begin{tabular} {c c c c c c c}
$\gamma_0$ & $\gamma_1$ & 0 & $\hdots$ & 0 & 0 & 0 \\
0 & $\gamma_0$ & $\gamma_1$ & $\hdots$ & 0 & 0 & 0 \\
0 & 0 & $\gamma_0$ & $\hdots$ & 0 & 0 & 0 \\
$\vdots$ & $\vdots$ & $\vdots$ & $\ddots$ & $\vdots$ & $\vdots$ & $\vdots$ \\
0 & 0 & 0 & $\hdots$ & $\gamma_0$ & $\gamma_1$ & 0 \\
0 & 0 & 0 & $\hdots$ & 0 & $\gamma_0$ & $\gamma_1$ \\
0 & 0 & 0 & $\hdots$ & 0 & 0 & $\gamma_0$ 
\end{tabular}
\right)
\label{eq-h-N1xN1}
\end{equation}
with $\gamma_0 ({\bf k}_{\parallel})=-[(\alpha_x+\beta_x)
+(\alpha_y+\beta_y)e^{-ik_2}
+(\alpha_z+\beta_z)e^{-ik_3}]$
and 
$\gamma_1 ({\bf k}_{\parallel})=
-[(\alpha_x-\beta_x)
+(\alpha_y-\beta_y)e^{i k_2}
+(\alpha_z-\beta_z)e^{i k_3}]$.

As discussed in the main text,
we also consider the 3D SSH systems 
with the open surfaces 
parallel to ${\bf a}_2$ and ${\bf a}_3$ 
and the half unit cells on one surface,
which have the thickness 
$L_1=2N_1$+1 and
$N_1$+1 of $A$ sites 
and $N_1$ of $B$ sites in ${\bf a}_1$ direction.
Such systems are described by the Hamiltonian given by    
\begin{align}
&\hat{H}_0=\sum_{{\bf k}_{\parallel}, \sigma=\uparrow,\downarrow} 
\left(
\begin{array}{ c }
C^{\dagger}_{{\bf k}_{\parallel} A \sigma} \\
C^{\dagger}_{{\bf k}_{\parallel} B \sigma}
\end{array}
\right)^T
\nonumber \\
&\left(
\begin{array}{cc}
0 &
h^{\dagger}_{(N_1+1)\times N_1}({\bf k}_{\parallel}) \\
h_{N_1 \times (N_1+1)} ({\bf k}_{\parallel}) &
0
\end{array}
\right)
\left(
\begin{array}{c}
C_{{\bf k}_{\parallel} A \sigma} \\
C_{{\bf k}_{\parallel} B \sigma}
\end{array}
\right),
\label{eq-H0-surface-half-unit-cells} 
\tag{S5}
\end{align}
where
$C^{\dagger}_{{\bf k}_{\parallel} A \sigma}=
(c^{\dagger}_{1, {\bf k}_{\parallel}  A \sigma}, 
c^{\dagger}_{2, {\bf k}_{\parallel}, A \sigma}, ..., 
c^{\dagger}_{N_1, {\bf k}_{\parallel} A \sigma}, 
 c^{\dagger}_{N_1+1, {\bf k}_{\parallel} A \sigma})$, 
$C^{\dagger}_{{\bf k}_{\parallel} B \sigma}=
(c^{\dagger}_{1, {\bf k}_{\parallel}  B \sigma}, 
c^{\dagger}_{2, {\bf k}_{\parallel} B \sigma}, ..., 
c^{\dagger}_{N_1, {\bf k}_{\parallel} B \sigma})$,
and the $N_1 \times (N_1+1)$ matrix $h_{N_1\times (N_1+1)}({\bf k}_{\parallel})$ is
\begin{equation}
\tag{S6}
h_{N_1\times (N_1+1)}({\bf k}_{\parallel})=
\left( 
\begin{tabular} {c c c c c c c c}
$\gamma_0$ & $\gamma_1$ & 0 & $\hdots$ & 0 & 0 & 0 & 0\\
0 & $\gamma_0$ & $\gamma_1$ & $\hdots$ & 0 & 0 & 0 & 0\\
0 & 0 & $\gamma_0$ & $\hdots$ & 0 & 0 & 0 & 0 \\
$\vdots$ & $\vdots$ & $\vdots$ & $\ddots$ & $\vdots$ & $\vdots$ & $\vdots$ & $\vdots$ \\
0 & 0 & 0 & $\hdots$ & $\gamma_0$ & $\gamma_1$ & 0 & 0 \\
0 & 0 & 0 & $\hdots$ & 0 & $\gamma_0$ & $\gamma_1$ &0 \\
0 & 0 & 0 & $\hdots$ & 0 & 0 & $\gamma_0$ & $\gamma_1$  
\end{tabular}
\right)
\end{equation}
with $\gamma_0 ({\bf k}_{\parallel})$
and 
$\gamma_1 ({\bf k}_{\parallel})$
identical to those in Eq.~(\ref{eq-h-N1xN1}).
The special case of $N_1$=1 
and $\beta_y$=$\beta_z$=0
results in 
two bulk bands and one absolutely flat midgap band given by
\begin{align}
\varepsilon_0({\bf k}_{\parallel} )=&0,  \tag{S7} \\
\varepsilon_{\pm}({\bf k}_{\parallel} )=&\pm \sqrt{2} 
\left[ \left( \alpha_x +\alpha_y \cos k_2 + \alpha_z \cos k_3  \right)^2 
 \right. \nonumber \\
& \left. +\beta_x^2 +\left( \alpha_y \sin k_2 + \alpha_z \sin k_3  \right)^2
\right]^{1/2}.  \tag{S8}
\end{align}
The 3D Hubbard SSH Hamiltonians with the open surfaces
are obtained by adding the diagonal on-site Hubbard terms 
to Eqs.~(\ref{eq-H0-surface-full-unit-cells}) and~(\ref{eq-H0-surface-half-unit-cells}),
which are analyzed through the Hartree-Fock approximation as presented in the main text. 

\section{Phase diagrams}

\begin{figure}
\renewcommand{\thefigure}{S1}
\includegraphics[width=\linewidth]{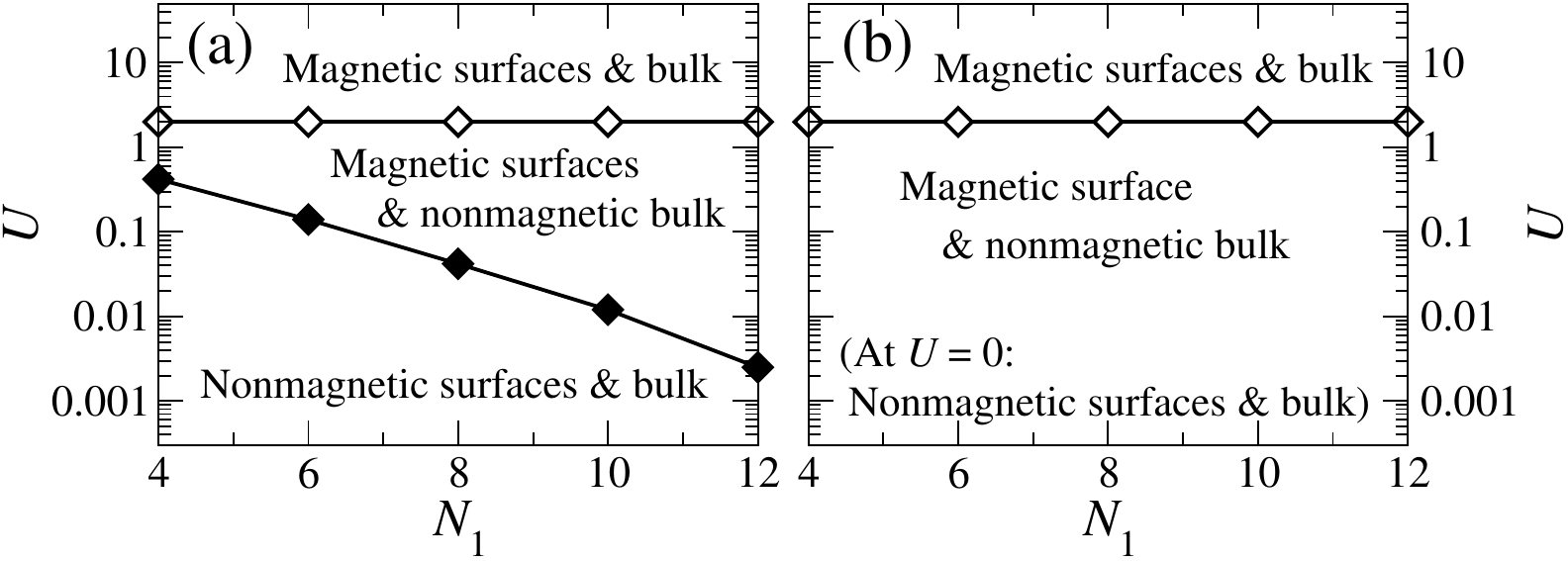}
\caption{
Phase diagrams in semilogarithmic $U$-$N_1$ planes,
(a) for the 3D topological Hubbard SSH systems 
with the full unit cells on both surfaces, 
and (b) for those with 
the half unit cells on one surface and full unit cells on the other.
$U$ represents the on-site Coulomb interaction and 
$N_1$ the number of the full unit cells in ${\bf a}_1$ direction. 
The parameter values are identical to those 
in Figs. 2(c), 3, and 4 in the main text,
except $U$ and $N_1$ varied here.
}
\label{fig-phase-diagram}
\end{figure}

The phase diagrams 
in the semilogarithmic $U$-$N_1$ plane
are shown 
in Fig.~\ref{fig-phase-diagram}(a)
for the 3D topological Hubbard SSH systems with the full unit cells on both surfaces 
and 
in Fig.~\ref{fig-phase-diagram}(b)
for those with the half unit cells on one surface,
which show the regions 
with the magnetic moments confined only on the topological surfaces. 
For the topological systems with full unit cells on both surfaces,
the critical Coulomb interaction for the surface spin polarization, $U_{\rm c,s}$, 
shown in solid symbols in Fig.~\ref{fig-phase-diagram}(a),
decays exponentially as $N_1$ increases,
reflecting the exponentially decaying surface band widths, as found for the 2D SSH systems~\cite{Qian20}.
In contrast, for the systems with half unit cells on one surface in Fig.~\ref{fig-phase-diagram}(b),
the surface spin polarization appears for any nonzero Coulomb interactions, 
as the surface band is absolutely flat.  
The critical Coulomb interaction for the intrinsic bulk spin polarization, $U_{\rm c,b}$,
shown in open symbols 
in Figs.~\ref{fig-phase-diagram}(a) and~\ref{fig-phase-diagram}(b),
remains independent of the system thickness, 
as the bulk band width is approximately unchanged.

\section{$N$D SSH Hamiltonian and topological analysis}
The SSH model is generalized to the arbitrary $N$-dimensional space.
Analogous to the 3D SSH system in 
%Fig.~\ref{fig-structure} 
Fig.~1
in the main text, 
we start with a $N$D hypercubic lattice,
and add uniform 
and staggered distortions
parameterized by $e_n$ and $d_n$, respectively,
in the $x_n$ direction, where $n$=1, 2, $\ldots$ , $N$.
Expressed in the Cartesian unit vectors,
primitive vectors are chosen as 
${\bf a}_1=2 (1+e_1) \hat{\bf x}_1$,
${\bf a}_2= - (1+e_1) \hat{\bf x}_1 + (1+e_2) \hat{\bf x}_2$, 
${\bf a}_3= - (1+e_1) \hat{\bf x}_1 + (1+e_3) \hat{\bf x}_3$, 
$\ldots$ ,
${\bf a}_N= - (1+e_1) \hat{\bf x}_1 + (1+e_N) \hat{\bf x}_N$.
The two-site basis is made of an $A$ site at 
$\sum_{n=1}^{N} d_n \hat{\bf x}_n$
and a $B$ site at $(1+e_1)\hat{\bf x}_1 -\sum_{n=1}^{N} d_n \hat{\bf x}_n$.

With $t_0$ representing the nearest neighbor hopping amplitude 
for the $N$D hypercubic lattice before the distortions are applied,
the Hamiltonian for the $N$D SSH system is given by 
\begin{align}
&\hat{H}_0=\sum_{\bf R} \left[
-(\alpha_1+\beta_1) c^{\dagger}_{{\bf R} B} c_{{\bf R} A}
-(\alpha_1-\beta_1) c^{\dagger}_{{\bf R} B} c_{{\bf R}+{\bf a}_1, A} \right.  \nonumber \\
&
\left. -\sum_{n=2}^{N}
(\alpha_n+\beta_n) c^{\dagger}_{{\bf R}+{\bf a}_n, B} c_{{\bf R} A} 
-(\alpha_n-\beta_n) c^{\dagger}_{{\bf R} B} c_{{\bf R}+{\bf a}_1+{\bf a}_n, A} \right] \nonumber \\
&+ H.c.,
\tag{S9}
\label{eq-H0-ND-real}
\end{align}
where $\alpha_n=t_0(1-e_n)$
and $\beta_n=2 t_0 d_n$
represent the linear modulation of the hopping parameters 
by the distortions and the spin indices are omitted.
Similar to the 3D SSH model, the Hamiltonian in reciprocal space is given by 
\begin{equation}
\tag{S10}
\hat{H}_0=\sum_{\bf k}
\left(
\begin{array}{ c c}
c^{\dagger}_{{\bf k} A} &
c^{\dagger}_{{\bf k} B}
\end{array}
\right)
\left(
\begin{array}{c c}
0 &
h^{*}({\bf k}) \\
h ({\bf k}) &
0
\end{array}
\right)
\left(
\begin{array}{c}
c_{{\bf k} A} \\
c_{{\bf k} B}
\end{array}
\right),
\end{equation}
where
${\bf k} = (k_1/2\pi) {\bf b}_1 + (k_2/2\pi) {\bf b}_2 + ... + (k_N/2\pi) {\bf b}_N$
with ${\bf b}_n$ ($n$=1, 2, ..., $N$) representing the primitive reciprocal vectors
and
$h({\bf k})=-[ (\alpha_1+\beta_1)
+(\alpha_1-\beta_1)e^{ik_1}]
-\sum_{n=2}^N 
[(\alpha_n+\beta_n) e^{-ik_n}
+(\alpha_n-\beta_n)e^{i(k_1+k_n)}]$.
The above Hamiltonian leads to the energy bands given by
\begin{align}
\varepsilon_{\pm}({\bf k})&=\pm 2 \left\{
\left[
\alpha_1 \cos{ \frac{k_1}{2} }
+\sum_{n=2}^{N}
\alpha_n \cos{ \left(\frac{k_1}{2}+k_n\right) }
\right]^2  \right.
\nonumber \\
+&
\left.
\left[
\beta_1 \sin{ \frac{k_1}{2} }
+\sum_{n=2}^{N}
\beta_n \sin{ \left(\frac{k_1}{2}+k_n\right) }
\right]^2
\right\}^{1/2}.
\tag{S11}
\end{align}
The gap opens and the system becomes an insulator for a half filling,
if the system is made of weakly coupled SSH chains, 
that is, if 
$\beta_n \neq 0$,
$\beta_{m \neq n} =0$,
and
$\alpha_n > \sum_{m \neq n} \alpha_m$
for $n$ = 1, 2, ..., or $N$.

The topological analysis is carried out  
in terms of the winding numbers, 
$\nu_1$, $\nu_2$, ..., $\nu_N$,
defined by
\begin{equation}
\tag{S12}
\nu_n=\frac{1}{2\pi i} \int^{2 \pi}_0 dk_n \frac{\partial}{\partial k_n} \ln h(k_1,k_2, ..., k_N),
\end{equation} 
with $n=1, 2, ..., N$.
Table~\ref{table-ND-winding} shows how the winding numbers depend on the parameters,
which reveals that the system is a topological insulator 
except when $\beta_1 > 0$, $\beta_{m \neq 1} = 0$  and 
$\alpha_1 > \sum_{m \neq 1} \alpha_m$.
As the nonzero winding numbers dictate the directions of the topological hypersurfaces, 
Table~\ref{table-ND-winding} also illustrates that there are two types of topological hypersurfaces.
One is the type of the hypersurfaces that are perpendicular to 
${\bf b}_1$, that is, parallel to ${\bf a}_2$,  ${\bf a}_3$, ..., and ${\bf a}_N$,
for the cases with $\nu_1=1$.
In this type, the open hypersurfaces have 
the $A$ sites only or the $B$ sites only,
as in the 3D SSH systems.
The other is the type of the hypersurfaces 
that are perpendicular to the SSH chains 
for the cases with  $\nu_{n\neq 1}=\pm 1$.
In this type, 
the open hypersurfaces would have the alternating $A$ and $B$ sites in 
a ($N-1$)-dimensional checkerboard pattern.  
These topological hypersurfaces would host 
nearly or absolutely flat, midgap hypersurface bands,
depending on whether the number of the hyperlayers is even or odd,
similar to the 3D SSH systems.  
Computational demonstrations of such flat bands for the $N$D topological SSH systems
with open hypersurfaces are left for future studies.

\begin{table*}
\renewcommand{\thetable}{S1}
\begin{tabular} {c | c | c c c c c c | c | c c c c c c | c}
\hline \hline
SSH chains & $\alpha_1$, $\alpha_2$, ... , $\alpha_N$ & $\beta_1$ & $\beta_2$ & ... & $\beta_n$ & ... & $\beta_N$ 
& $\nu_1$ & $\nu_2$ & $\nu_3$ & ... & $\nu_n$ & ... & $\nu_N$ & TI/nonTI  \\
\hline
\multirow{ 2}{*}{Along $\hat{x}_1$} & \multirow{ 2}{*}{$\alpha_1 > \sum_{m \neq 1} \alpha_m$}  
& $+$ & 0 & ... & 0 & ...& 0 & 
0 & 0 & 0 & ... & 0 & ... & 0 & nonTI \\
&  & $-$ & 0 & ... & 0 & ...& 0 & 
1 & 0 & 0 & ... & 0 & ... & 0 & TI \\
\hline
\multirow{ 2}{*}{Along $\hat{x}_2$} & \multirow{ 2}{*}{$\alpha_2 > \sum_{m \neq 2} \alpha_m$}  & 0 & $+$ & ... & 0 & ...& 0 & 
0 & $-1$ & 0 & ... & 0 & ... & 0 & TI \\
&  & 0 & $-$ & ... & 0 & ...& 0 & 
1 & 1 & 0 & ... & 0 & ... & 0 & TI \\
\hline
\vdots  & \vdots & \vdots & \vdots & \vdots & \vdots & \vdots & \vdots & \vdots & \vdots & \vdots & \vdots & \vdots & \vdots & \vdots & \vdots \\
\hline
\multirow{ 2}{*}{Along $\hat{x}_n$} & \multirow{ 2}{*}{$\alpha_n > \sum_{m \neq n} \alpha_m$}  & 0 & 0 & ... & $+$ & ...& 0 & 
0 & 0 & 0 & ... & $-1$ & ... & 0 & TI \\
&  & 0 & 0 & ... & $-$ & ...& 0 & 
1 & 0 & 0 & ... & $1$ & ... & 0 & TI \\
\hline
\vdots & \vdots & \vdots & \vdots & \vdots & \vdots & \vdots & \vdots & \vdots & \vdots & \vdots & \vdots & \vdots & \vdots & \vdots & \vdots \\
\hline
\multirow{ 2}{*}{Along $\hat{x}_N$} & \multirow{ 2}{*}{$\alpha_N > \sum_{m \neq N} \alpha_m$}  & 0 & 0 & ... & 0 & ...& $+$ & 
0 & 0 & 0 & ... & 0 & ... & $-1$ & TI \\
&  & 0 & 0 & ... & 0 & ...& $-$ & 
1 & 0 & 0 & ... & 0 & ... & 1 & TI \\
\hline
\end{tabular}
\caption{
Winding numbers $\nu_n$ with $n$ = 1, 2, ..., $N$
for the hypersurfaces parallel to 
all ${\bf a}_m$'s with $m \neq n$
for the $N$D SSH systems.
The parameters $\alpha_n$ and $\beta_n$ with $n$ = 1, 2, ..., $N$
are the parameters of the $N$D SSH Hamiltonian in Eq.~(\ref{eq-H0-ND-real}),
associated with the distortions.
TI and nonTI mean topological and nontopological insulators, respectively.   
}
\label{table-ND-winding}
\end{table*}